\documentclass[prb,aps,twocolumn,floatfix]{revtex4}
\usepackage[dvips]{graphicx}
\usepackage{latexsym}
\usepackage{amsmath,amsfonts,color}

\newcommand{\vect}[1]{\boldsymbol{#1}}

\renewcommand{\(}{\left(}
\renewcommand{\)}{\right)}

\begin{document}
\title{Spin and pair density wave glasses}
 \author{David F. Mross}
\affiliation{Department of Physics and Institute for Quantum Information and Matter, California Institute of Technology,
Pasadena, CA 91125, USA}
 \author{T. Senthil}
\affiliation{Department of Physics, Massachusetts Institute of Technology,
Cambridge, MA 02139, USA }
\begin{abstract}

Spontaneous breaking of translational symmetry---known as `density wave' order---is common in nature. However such states are strongly sensitive to impurities or other forms of frozen disorder leading to fascinating glassy phenomena. We analyze impurity effects on a particularly ubiquitous form of broken translation symmetry in solids: a Spin Density Wave (SDW) with spatially modulated magnetic order. Related phenomena occur in Pair Density Wave (PDW) superconductors where the superconducting order is spatially modulated. For weak disorder, we find that the SDW / PDW order can generically give way to a SDW / PDW glass---new phases of matter with a number of striking properties, which we introduce and characterize here. In particular, they exhibit an interesting combination of conventional (symmetry-breaking) and spin glass (Edwards-Anderson) order. This is reflected in the dynamic response of such a system, which---as expected for a glass---is extremely slow in certain variables, but---surprisingly---is fast in others. Our results apply to all uniaxial metallic SDW systems where the ordering vector is incommensurate with the crystalline lattice. In addition, the possibility of a PDW glass has important consequences for some recent theoretical and experimental work on $La_{2-x}Ba_xCu_2O_4$.
\end{abstract}
\maketitle

\section{Introduction}
A variety of electronic solids settle into equilibrium states that spontaneously break the translational symmetry of the underlying crystal \cite{monceau}. Well known examples are Charge Density Wave (CDW) and Spin Density Wave (SDW) orders where either the electron's charge or its spin forms a frozen periodically oscillating pattern. Density wave order has been found in conventional metals as well as in strongly correlated systems such as the underdoped cuprates, iron pnictides, and organic materials, and are intertwined with many other fascinating phenomena such as for instance high temperature superconductivity.

It has long been recognized that density wave orders of various kinds are strongly sensitive to the presence of impurities. There is a large literature on the fascinating effects of quenched disorder on charge density wave systems. 
In contrast, despite the common occurrence of spin density wave ordering, surprisingly little attention has been devoted to impurity effects on SDW systems, and this is the subject of this paper. 

\begin{figure}[ht]
\includegraphics[width=0.8\columnwidth]{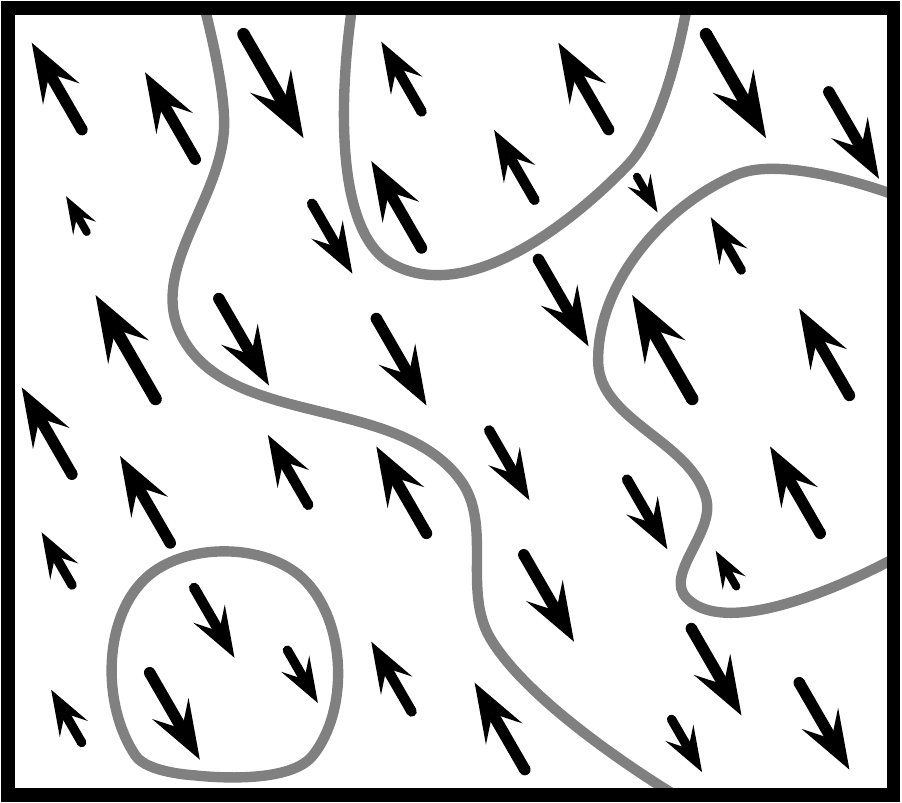}
\caption{Magnetic moments in the ground state of the SDW glass. The disorder pins the domain walls into a random configuration, but the structure of `anti-phase' domain walls persists. Upon crossing of each domain wall, the local magnetization changes sign. Thus the disordered state inherits the collinear structure from the parent SDW, where the axis along which the moments point is selected sponteneously, breaking the spin rotation symmetry. }
\label{fig:sdw1}
\end{figure}

\begin{figure}[ht]
\includegraphics[width=0.8\columnwidth]{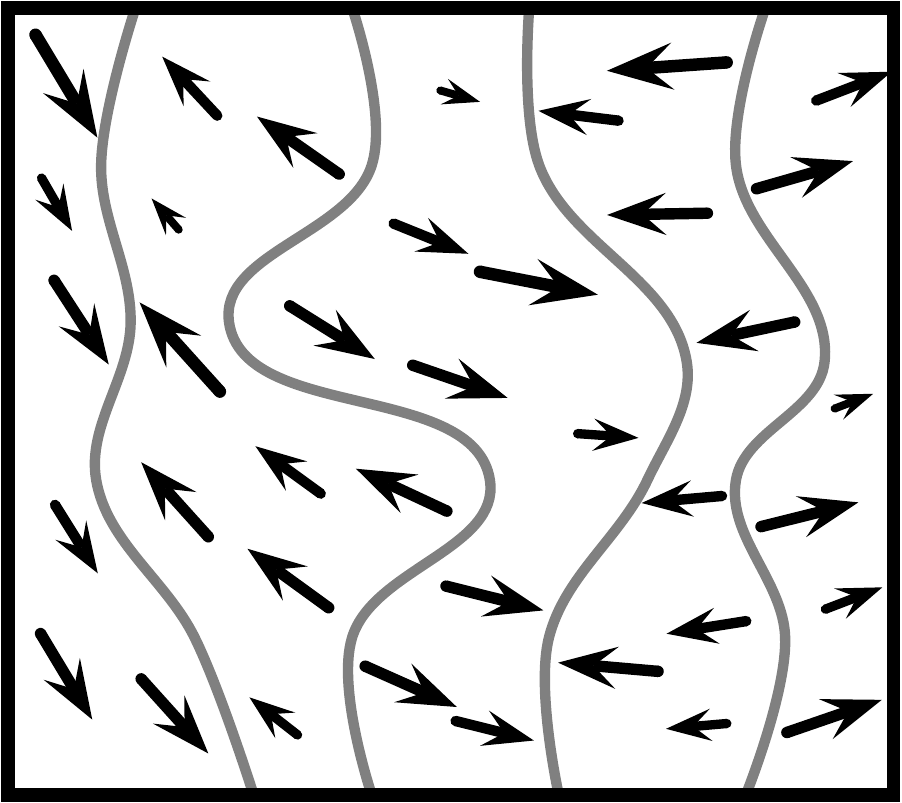}
\caption{Goldstone mode in the SDW glass. The configuration of the domain walls is identical to the ground state, but the orientation of the moments slowly varies in space}
\label{fig:sdw2}
\end{figure}

It is important to distinguish between \emph{collinear} SDWs where the spin orientation oscillates in space along a fixed common direction and \emph{spiral} SDWs where the spin rotates around an axis as a function of space while the magnitude $|\vec S(\vect r)|$ is constant. Both kinds of SDW order break both spin-rotation and lattice translation symmetries, but the latter retains a combination of the two as a symmetry. As a consequence, disorder effects on collinear SDWs are stronger and are the focus of our study. 
We show that weak non-magnetic disorder transforms the SDW state into a new glassy state of matter---distinct from the conventional spin glass---which we dub the SDW glass (see Fig. \ref{fig:sdw1}). In addition to the characteristics of a conventional spin glass, i.e. the presence of locally frozen moments but absence of long-range spin order, the SDW glass spontaneously breaks spin-rotation symmetry and hosts an associated Goldstone mode (see Fig. \ref{fig:sdw2}).

Questions closely related to the ones we study arise in considering the effects of impurities on Pair Density Wave (PDW) superconductors of the kind proposed to be realized in the high temperature superconductor $La_{2-x}Ba_xCu_2O_4$ (see Ref.~\onlinecite{fktrev} for a recent review). These are superconducting states where the pair amplitude is modulated in space. Such states were originally conceived by Larkin and Ovchinnikov\cite{LO} when considering the effects of a Zeeman magnetic field on an s-wave superconductor (closely related states were considered by Fulde and Ferrell\cite{FF}). Motivated by the phenomenology observed in $La_{2-x}Ba_xCu_2O_4$, Refs.~\onlinecite{bfk,bfkt} proposed that a PDW state is realized at zero magnetic field in this system. Its microscopic origin remains to be understood. Like the analogous SDW order, the PDW state is also expected to be strongly sensitive to impurities. Refs.~\onlinecite{bfk,bfkt} suggested that disorder necessarily introduces half-quantum ($hc/4e$) vortices leading to broken time reversal symmetry. We revisit this issue and find 
that in the phase analogous to the SDW glass, such vortices will not be induced by weak disorder and that time reversal symmetry is preserved. 
A fluctuating version of the PDW state also appears as a `mother' state that controls the 
physics of the pseudogap regime in the cuprates in a recently proposed theory \cite{amperean}. 

\section{SDW and PDW: Order parameters and topological defects}
\label{optd}
For simplicity we focus initially on a unidirectional collinear SDW at wave-vector $\vect Q$ in which the spin oscillates as 
\begin{align}
 \vec S(\vect r) = S_0\cos \(\vect Q \cdot \vect r\)\vec N\label{eq:sdw}, 
\end{align}
where $\vec N$ is a real unit vector and $S_0$ is the amplitude. We further assume that $\vect Q$ is incommensurate with the crystalline lattice. Such SDW order is sometimes also referred to as a `spin stripe'. A simple Landau argument shows that this pattern of spin ordering induces CDW order at wave vector $2\vect Q$: 
 \begin{align}
 \rho_\text{CDW} \sim \rho^0_{2\vect Q} \cos \(2\vect Q \cdot \vect r\)\label{eq:scdw}
\end{align}

To discuss situations in which the SDW order is fluctuating (either due to quenched disorder or due to thermal/quantum fluctuations) we write the spin as
\begin{equation}
\vec S(\vect r) \approx S_0 \cos(\vect Q \cdot \vect r + \theta(\vect r) )\vec N.
\end{equation}
$\theta$ describes the phase of the SDW (or the stripe displacement, in the stripe picture). We take both $\theta$ and $\vec N$ to be space (and possibly time) dependent but varying on length scales long compared to the SDW wavelength. The fluctuating SDW order parameter near wave vector $\pm \vect Q$ is thus 
\begin{equation}
\vec S_{\pm \vect Q}(\vect x) =S_0 e^{\pm i\theta(\vect x) } \vec N(\vect x).
\end{equation}
 The corresponding fluctuating CDW order parameter is 
 \begin{equation}
 \rho_{\pm 2\vect{Q}} = \rho^0_{2\vect Q} e^{\pm 2i\theta}.
 \end{equation} 
 It is useful to formulate discussions of fluctuations in terms of the two separate fields $b \equiv e^{i\theta}$ and $\vec N$. Both the SDW order parameter $\vec S_{\vect Q} \sim b \vec N$ and the CDW order parameter $\rho_{2 \vect Q} \sim b^2$ are composites made out of $b$ and $\vec N$. Clearly the $b, \vec N$ representation has a $\mathbb{Z}_2$ gauge redundancy under $b \rightarrow -b, \vec N \rightarrow -\vec N$ so that neither of them are directly physical \cite{zaanen,ks,ssmor,short,long}. The SDW/CDW order parameters are of course gauge invariant. 

Let us now turn to the closely analogous PDW state (sometimes called a striped superconductor or the Larkin-Ovchinnikov phase). This is a superconductor in which the pair wave function $\Delta$ is modulated in space: 
\begin{equation}
\Delta(\vect r) = \Delta_0 \cos(\vect Q \cdot \vect r)
\end{equation}
This too will induce CDW order at wave vector $2\vect Q$. When fluctuating we may write 
\begin{equation}
\Delta(\vect r) \approx \Delta_0 \cos(\vect Q \cdot \vect r + \theta (\vect r))e^{i\phi(\vect r)}
\end{equation}
The Fourier components of $\Delta(\vect r)$ near $\pm \vect Q$ are thus $\sim e^{i(\phi \pm \theta)}$. There is again a $\mathbb{Z}_2$ gauge redundancy under $\theta \rightarrow \theta + \pi, \phi \rightarrow \phi + \pi$. 

The PDW state is thus conceptually very similar to a SDW state with just $XY$ spin symmetry. However there is an important difference in the action of time reversal symmetry. The PDW state preserves time reversal while the SDW breaks it. Formally this is because the $U(1)$ charge conservation symmetry (broken in the SC) does not commute with time reversal while spin rotations do. Nevertheless we will consider both orders within the same framework. 
Unless otherwise specified we will phrase our discussion in terms of SDW order. 

The structure of topological defects in these density wave states\cite{zaanen,ks,ssmor,short,long} will play a crucial role below. Of particular importance are dislocations in the CDW pattern. These are line defects in 3D and point defects in 2D around which $\theta$ winds. The `elementary' strength-$1$ dislocation where the CDW phase $2\theta$ winds by $2\pi$ requires that $\vec N$ winds by $\pi$ so that the SDW order $e^{i\theta}\vec N$ is single valued. In contrast strength-$2$ dislocations have $2\theta$ wind by $4\pi$ without any winding of $\vec N$. 

Exactly the same considerations also apply in the superconducting context as described in Refs.~\onlinecite{bfk,agterberg,leo}. It is interesting to consider the physical interpretation of the various topological defects in this case. The strength-$1$ CDW dislocation now requires that the superconducting phase $\phi$ wind by $\pi$. This corresponds to a superconducting vortex with magnetic flux $\frac{hc}{4e}$, {\em i.e} half the usual flux quantum. Strength-$2$ CDW dislocations in contrast do not bind to superconducting vortices. 

 \section{Impurities: Models and preliminaries}
We want to consider the fate of the SDW in the presence of weak non-magnetic impurities. Such impurities lead to a random potential $V(\vect x)$ that couples linearly to the CDW order parameter, i.e.
\begin{align}
H_\text{dis.} = \vect F(\vect x) \cdot \vect \nabla \theta + V_{2 \vect Q}^*(\vect x) e^{2i \theta} + V_{2 \vect Q}(\vect x) e^{-2i \theta}\label{eq:dis},
\end{align}
The first term couples to the long wavelength part of the charge density (with $\vect F$ random) and the second to the density near the ordering wave vector. Here $V_{2 \vect Q}(\vect x) = \int_{\vect q \approx 2 \vect Q} e^{i \vect q \cdot \vect x}V(\vect q)$. There is however no linear coupling to the primary SDW order parameter. 
The impurity coupling is captured by a simple lattice model: 
\begin{align}
H =& - \frac{J}{2} \sum_{<ij>} \vec S_{\vect Q i}^* \cdot \vec S_{\vect Q j } e^{i\eta_{ij}} - \frac{v}{2}\sum_i\rho_{2 \vect Q _i}e^{-i \alpha_i} + c.c \nonumber \\
 = & -J\sum_{<ij>} \vec N_i \cdot \vec N_j \cos(\theta_i - \theta_j + \eta_{ij}) \nonumber \\
& - v\sum_i \cos(2\theta_i - \alpha_i) \label{eqn:firstmodel}
\end{align}
Here $\alpha_i, \eta_{ij}$ are random uncorrelated variables. Other equivalent lattice models may be formulated and are described in the Appendix \ref{app.model}.

As is well known\cite{larkin,imryma} the `random field' disorder destroys LRO in the CDW for physical dimensions $d < 4$. The elastic energy cost of adjusting to disorder over a scale $L$ scales as $L^{d-2}$ while the energy gain due to the disorder potential scales as $L^{d/2}$, thus the latter dominates for $d<4$. As a consequence, beyond a length scale (known as the Larkin length) $\xi_L \sim (J/v)^{2/(4-d)}$ long range CDW order is destroyed. This immediately implies the absence of long range SDW order as well (as long range SDW order if present would have induced CDW order). At distances longer than $\xi_L$ the disordered SDW enters a phase of matter that we dub the SDW glass and whose physics we describe below. 

\section{SDW glasses in 3d}\label{sec.3d}
We begin our analysis in $d=3$ dimensions by reviewing the physics of pinned CDW systems. In pioneering work, Ref.~\onlinecite{GLD94} proposed that at weak disorder the pinned CDW enters an `elastic glass' phase where long dislocation loops do not occur. This has been substantiated by numerical calculations\cite{huse} and by general scaling arguments\cite{fisher}. The resulting state is described by a random field XY model for the CDW order parameter where dislocations are suppressed. Many approximate treatments, notably a sophisticated Functional Renormalization Group (FRG) calculation\cite{fish,GLD94}, show that the CDW order parameter develops {\em power law} correlations:
\begin{equation}
\overline{\rho_{2\vect Q}^*(\vect x)\rho_{2\vect Q}(\vect x')} \sim \frac{1}{|\vect x - \vect x'|^{d_c}}
\end{equation}
The exponent $d_c$ is universal. To leading order in the $\epsilon$ expansion, $d_c = \frac{\pi^2\epsilon}{9}$. Thus for $d = 3$, $d_c$ is estimated to be $\approx 1.1$. The power law decay of the spatial CDW correlations implies power-law Bragg peaks in the CDW structure factor
 \begin{align}
 S_\text{CDW}(2\vect Q+ \delta \vect q) \sim |\delta \vect q|^{d_c -3},
 \end{align}
The elastic glass phase is therefore also known as the `Bragg glass' phase. 

Let us now consider the implications for the SDW order. The impurities do not couple linearly to $\vec N$ but will lead to random exchange energies (`random bond disorder'). However the absence of dislocations in $\theta$ means that there is no frustration of the collinear ordering of $\vec N$. Thus at weak disorder $\vec N$ will continue to have true long range order. The SDW order will has power law correlations inherited from the correlations of $e^{i\theta}$. 
\begin{eqnarray}
\overline{\vec S_{\bf Q}^*({\bf x}) \cdot
\vec S_{\bf Q}({\bf x'})} & = & \overline{e^{i\theta({\bf x})} e^{-i\theta({\bf x'})} \vec N({\bf x}) \cdot \vec N({\bf x'})} \nonumber \\
& \sim & \overline{e^{i\theta({\bf x})} e^{-i\theta({\bf x'})}} \nonumber \\
& \sim & \frac{1}{|{\bf x} - {\bf x'}|^{d_s}} \nonumber
\end{eqnarray}
The exponent $d_s$ can be estimated within the FRG in $d = 4-\epsilon$. At leading order in the $\epsilon$ expansion we have $d_s = \frac{\pi^2 \epsilon}{36}$. This is $\frac{1}{4}$ of the CDW exponent $d_c$, as the probability distribution for $\theta$ is Gaussian to this order (for a recent discussion see Ref.~\onlinecite{fedetal14}). This gives the estimate $d_s \approx 0.27$ in $d = 3$. However beyond leading order the distribution will not be Gaussian\cite{fedetal14}, and hence in general $d_s \neq \frac{d_c}{4}$. Note that the SDW correlations decay much slower than the CDW correlations. This is expected since the CDW order is the one directly affected by the disorder.

 Correspondingly, the spin structure factor exhibits power law Bragg peaks
 \begin{align}
 S_\text{SDW}(\vect Q+ \delta \vect q) \sim |\delta \vect q|^{d_s-3}.
 \label{3dsdwgsfc}
 \end{align}
This power law Bragg peak should be visible in neutron diffraction measurements on weakly disordered SDW materials. 

But what does it mean for $\vec N$ to be ordered? $\vec N$ is not gauge invariant and hence not directly observable. However ordering of $\vec N$ implies ordering of the spin quadrupole moment 
\begin{equation}
Q_{\alpha\beta} = N_\alpha N_\beta - \frac{\delta_{\alpha \beta}}{3} \vec N^2
\end{equation}
Thus even though the SDW order is destroyed long range spin quadrupole order (also known as spin nematic order) is preserved. The system develops spontaneous spin anisotropy without long range SDW ordering. 

This SDW glass phase has a simple physical description. The spins are frozen in time but the phase of the SDW is randomly disordered in space. The spin nematic order means that the spins retain a common axis along which they randomly point up or down. The freezing of the spins means that there is a non-zero Edwards-Anderson spin glass order parameter 
\begin{align}
q_\text{EA}\equiv&\lim \limits_{t \rightarrow \infty} \overline{\langle \vec S(\vect x,0)\cdot \vec S(\vect x,t) \rangle}\\
=& \lim \limits_{t \rightarrow \infty} \langle \cos \(\theta(\vect x,t)-\theta(\vect x,0)\)\rangle \langle \vec N \rangle^2\neq 0,
\end{align}
The disordered SDW is thus a uni-axial spin glass in a Heisenberg spin system with the axis of spin orientation determined spontaneously. It is clearly distinct from the conventional Heisenberg spin glass.

The spin nematic order in the SDW glass phase leads to propagating Goldstone modes (nematic director waves). The structure of magnetic moments characteristic for the ground state and for a soft excitation are shown in Figs. \ref{fig:sdw1},\ref{fig:sdw2}. This should be contrasted with the Halperin-Saslow\cite{halperinsaslow} spin wave modes in a Heisenberg spin glass which are typically damped. 

 \begin{figure}[ht]
\includegraphics[width=0.8\columnwidth]{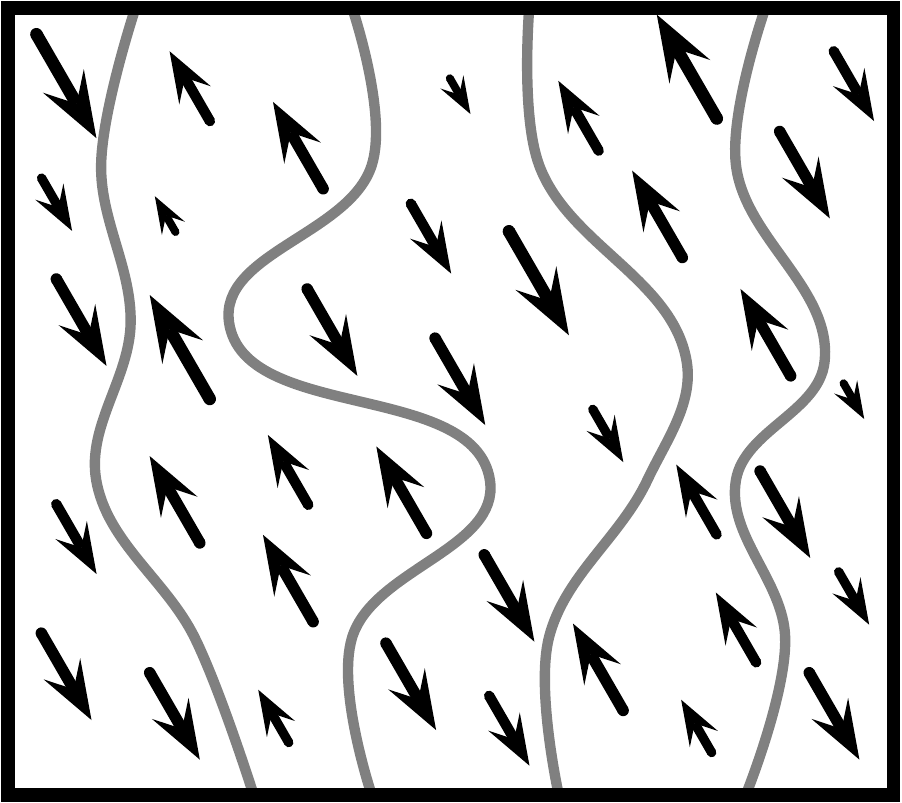}
\caption{In the SDW Bragg glass, the anti-phase domain walls are pinned in a configuration without dislocations.}
\label{fig:sdw3}
\end{figure}

The SDW glass exhibits striking differences from conventional spin glasses when placed in a weak magnetic field. Specifically, we will contrast the SDW glass to a Heisenberg spin glass (HSG) given by
\begin{align}
 H_\text{HSG} = \sum_{\vect x \vect y}J_{\vect x,\vect y}\vec S_{\vect x} \cdot \vec S_{\vect y},
\end{align}
where $J_{\vect x,\vect y}$ are random exchange couplings. Recall that in the absence of disorder, a collinear SDW aligns itself perpendicular to the applied field, with a slight canting of the moments, like an antiferromagnet. In the isotropic SDW glass, the canting is achieved by a rotation of $\vec N$. This corresponds to a fast (Goldstone) mode and therefore \emph{does not} exhibit slow (glassy) dynamics. In the HSG such a fast mode does not exist and the time scales for adjusting to the field are necessarily long.\\
For an anisotropic SDW glass $\vec N$ remains pinned to a specific direction in a weak enough magnetic field, and the system can only respond via the slow dynamics of $e^{i \theta}$. The cross-over value of the magnetic field $B_c$ between fast and slow dynamics is thus given by the strength of the anisotropy. In the HSG, a similar cross-over occurs, but at a much larger scale, determined by the typical exchange coupling $J$.\\

The same analysis presented here for the SDW also applies for a PDW. The analog of the spin nematic order parameter is a uniform charge-4$e$ superconducting order $\Delta_{4e}$. However we are not aware of any system that is proposed to host a PDW at zero magnetic field in $d=3$ dimensions. Our results should be pertinent though to Larkin-Ovchinikov pairing induced by a magnetic field in three dimensional superconductors. The breaking of time reversal allows additional terms in the Hamiltonian, in particular a linear coupling between the gradients of the phases of CDW and the condensate, i.e. in the continuum limit
\begin{align}
H_B = \lambda \int_{\vect x} \vect \nabla \phi \cdot \vect \nabla \theta 
\end{align}
 However, the PDW glass phase is perturbatively stable against such a term. Clearly the ground state in the absence of disorder is unaffected by $H_B$ for small $\lambda$. In the presence of disorder, at long distances $L \gg \xi_L$, $H_B$ contributes to the random bond energy for the charge-4$e$ superconducting order parameter $\Delta_{4e}$, which is irrelevant in the PDW glass where $\langle \Delta_{4e}\rangle\neq 0$. The same conclusion can be readily obtained through an FRG analysis (see Appendix \ref{app.fflofrg}).

\section{SDW and PDW glasses in 2d}

We now turn to $d=2$. Once again the random field will destroy long range CDW and hence long range SDW order. The fate of dislocations is however more subtle. 
In the simpler problem of the 2D random field XY model (appropriate to describe unidirectional CDW ordering not derived from a more primary SDW or PDW order), topological defects always proliferate at long scales, leading to exponentially decaying correlations. These defects cost elastic energy which must be balanced against the energy gain due to the random correlated potential induced by the random field. Refs.~\onlinecite{zlf99,lg00} show that at long enough scales the optimized potential energy for introducing vortices dominates so that it is always favorable to nucleate defects. 

In the SDW system, single and doubled CDW dislocations have different elastic cost---the energy of the former depends on the stiffness associated with spin distortions while the energy of the latter does not. Hence they could potentially behave very differently. While one expects that doubled dislocations are always generated at long length scales the fate of isolated single dislocations is less clear. 

Let us first describe a putative state where doubled dislocations have proliferated but single ones have not. In such a state the CDW correlations, and hence the SDW correlations decay exponentially. Despite that there is long range spin nematic order. Thus this is a 2D SDW glass phase with coexisting spin nematic order.

It is particularly interesting to note the meaning of these issues in the PDW context. The analog of the spin nematic then is a uniform charge-4$e$ superconductor. Further (see Section \ref{optd}) a strength-$1$ CDW dislocation is bound to a $\pi$ winding of the SC phase. This corresponds to a half-quantum vortex with $\frac{hc}{4e}$ flux. Clearly two different dislocations of this kind are possible depending on whether the flux is positive or negative. Any such single dislocation necessarily breaks time reversal symmetry. Thus if disorder nucleates these single dislocations then the PDW glass will break time reversal spontaneously by generating randomly placed $\pm \frac{hc}{4e}$ half-quantum vortices of either sign. Such vortices can be imaged using local probes of magnetism such as a scanning SQUID microscope, and can serve as a key experimental test of the proposed PDW state in $La_{2-x}Ba_xCu_2O_4$.

We therefore now pose the question of whether such single dislocations are {\em necessarily} generated at {\em weak} disorder in either the SDW or PDW system.

For simplicity and to provide a unified discussion of both SDW and PDW systems, we will specialize to XY spins. The fate of dislocations may be discussed within an appropriate elastic model which takes the form 
 \begin{align}
 H = &\int d^2 x \frac{K_s}{2} (\vect \nabla \phi)^2+\frac{K_c}{2} (\vect \nabla \theta - \vect f)^2 - v \cos \(2 \theta - \alpha\)\label{eq.h2d}
\end{align}
Here $K_c, K_s$ are the stiffnesses of the CDW and SDW (or PDW) order parameters respectively. The phase of the SDW (or PDW) order parameter is $\phi \pm \theta$, and that of the CDW order parameter is $2\theta$. The disorder is taken to be delta-correlated $\overline{f_i(\vect x) f_j(\vect y)} = D_F \delta_{ij}\delta(\vect x - \vect y)$, $\overline{\alpha(\vect x) \alpha(\vect y)}=\delta(\vect x - \vect y)$. Eq. (\ref{eq.h2d}) is the basis for our analysis of dislocations in the rest of this section.

The $\theta$ sector is again described by an $XY$ model with random anisotropy. However, in this case there is no stable dislocation free phase even for weak disorder. We will begin with discussions by reviewing some basic facts on this model:
In the case of random forces ({i.e} the $\vect f$ term), but without random fields, the relative displacement $\theta_{\vect x}-\theta_{\vect x'} $ grows logarithmically with distance, with a coefficient given by the variance $D_F$ of the random force \cite{fisher}. The energy cost of the cheapest dislocation is\cite{natter} $E \sim K_c(1 - \sqrt{D_F/D_F^c})\log L$, i.e. there is a critical strength $D_F^c$ of the random force, below which no isolated dislocations are present in the ground state.
In the presence of random fields but with dislocations excluded, $D_F$ is renormalized without bound\cite{co} as $D_F = C \log L$. The coefficient $C$ is temperature dependent. This implies
\begin{align}
 \overline{\langle \theta_{\vect x} - \theta_{\vect x'} \rangle^2} \sim \log^2|\vect x - \vect x'|.
\end{align}
Thus $C_{n}(\vect r)= \overline{\left\langle e^{i n\theta({\bf x + \vect r})} e^{-i n \theta({\bf x})}\right \rangle}$ decays faster than any power law, and the associated Bragg peaks are indistinguishable from the case of short-range correlations. 
When dislocations are allowed, they always become relevant\cite{fisher,lg00} for large enough $L$ since $D_F(L)\sim \log L > D_F^c$. Beyond the scale where dislocations proliferate, $C_{n}(\vect r)$ decays exponentially. 

\begin{figure}[ht]
\includegraphics[width=0.8\columnwidth]{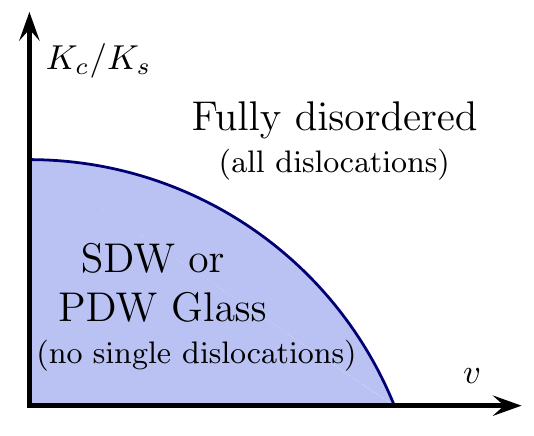}
\caption{Schematic phase diagram for a disordered SDW (or PDW) in 2D as a function of disorder strength $v$ and the ratio $K_s/K_c$ of the stiffnesses associated with spin waves / phonons in the absence of disorder}
\label{fig:phases}
\end{figure}

\begin{figure*}[ht]
 \includegraphics[width=17cm]{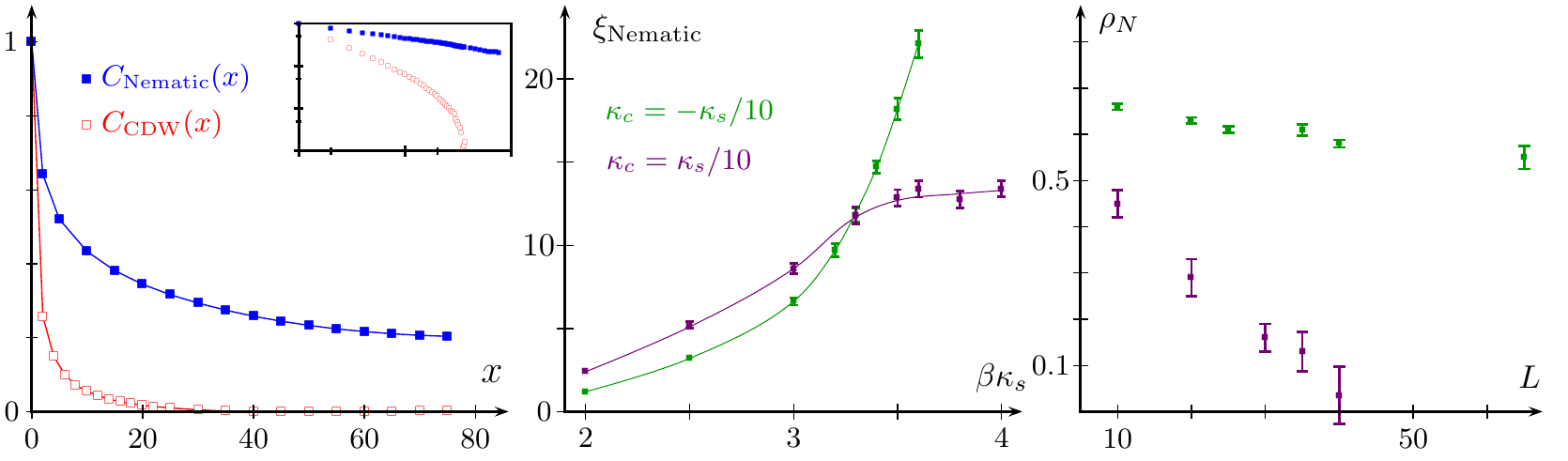}
\caption{Summary of numerical data. Left: Spatial decay of the correlations functions $C_\text{Nematic}$ and $C_\text{CDW}$. The inset shows a $\log-\log$ plot of the same data. Combined error due to sampling and disorder averaging is below the size of the shown data points. The qualitative feature that nematic correlations decay much slower than CDW correlations is generic for low temperatures and weak disorder. Center: Temperature dependence of the spin-nematic correlation length $\xi_\text{Nematic}$ shown for parameters corresponding to a SDW glass phase (green) or a fully disorderd phase (purple) at low temperatures. For the first set of parameters, $\xi_\text{Nematic}$ grows rapidly as temperature is lowered, indicating a phase transition into the SDW glass phase. For the second set of parameters, $\xi_N$ shows no sign of divergence and is expected to saturate at a finite value such that the low temperature glassy phase is smoothly connected to the high temperature phase. Solid lines are drawn as a guide to the eye. Right: The stiffness $\rho_N$ for spin nematic fluctuations, as a function of system size shown for both the SDW glass and the fully disordered phase. In the SDW glass ($\kappa_c = -0.1 \kappa_s$), $\rho_N$ depends only weakly on system size $L$ and always takes a finite value, while for $\kappa_c = 0.1 \kappa_s$ it exhibits substantial sample dependence and rapidly decays with system size.}\label{fig.bigfig}
\end{figure*}

The analysis in Appendix \ref{app.scales} shows that single dislocations proliferate at a scale 
\begin{equation}
\xi_V^{(1)} \approx 
\xi_L e^{\sqrt{\frac{\pi}{16 C }( 1 +
 K_s/K_c )^2\log \xi_L/a} } 
\end{equation}
Doubled dislocations on the other hand proliferate at a different length scale 
\begin{equation}
\xi_V^{(2)} \approx
\xi_L e^{\sqrt{\frac{\pi}{4 C }\log \xi_L/a}} 
\end{equation}
It follows that for large $\frac{K_s}{K_c}$ strength-$2$ dislocations proliferate at a shorter length scale. In that case the physics for distances longer than $\xi_V^{(2)}$ is modified from that described by the elastic model. The prevalence of doubled dislocations renormalizes $K_c$ to zero without affecting $K_s$. The net energy cost of the spin distortion associated with a strength-$1$ dislocation is then $\frac{\pi}{4} K_s \ln (\frac{\xi_L}{a})$ but now the optimal potential energy gain of the dislocation will just be a constant (see Appendix \ref{app.gauge}). Thus in this case strength-$1$ dislocations are suppressed even at the longest length scales. In the SDW glass this implies that spin nematic long range order survives at weak disorder in this regime. In the PDW glass in this regime time reversal is preserved and the disorder does not introduce half-quantum vortices. Rather there is true long range order in the charge-$4e$ superconducting order parameter. 

In the opposite regime of large $\frac{K_c}{K_s}$ single dislocations will proliferate first even at weak disorder. In either regime at strong disorder single dislocations will proliferate. 
The schematic phase diagram is shown in Fig. \ref{fig:phases}.

\subsection{Numerics}

To support the arguments presented above we performed Monte-Carlo simulations of the Hamiltonian
\small
\begin{align}
H_\text{MC}=&-\kappa_c\sum_{\langle ij\rangle} \cos (2\theta_{i} - 2\theta_j+2\eta_{ij}) \nonumber\\
& - \kappa_s \sum_{\langle ij \rangle}\cos (\theta_i - \theta_j+\eta_{ij})\cos (\phi_i - \phi_j)\nonumber\\
& +v\sum_{i} \cos \(2\theta_i + \alpha_i)\)\label{eq:num}.
\end{align}
\normalsize
where $\langle\vect r,\vect r'\rangle$ denotes nearest neighbors on a square lattice, $\alpha(\vect r)$ is a uniform random variable for each site and $\eta_{r,r'}$ is a gaussian random variable for each link with variance $D_F$. We note that $\eta_{r,r'}$ is generated under renormalization and does not need to be explicitly added to the Hamiltonian. However, it is convenient for studying the case of weak random field. There, the large difference between $\xi_L$ and $\xi_V$ makes numerical simulations challenging. Taking $D_F \lesssim D_F^c$ significantly reduces $\xi_V$ while otherwise retaining the same physics as the $D_F=0$ model.

Configurations were generated using the standard Hybrid Monte Carlo algorithm\cite{hmc}. The correlations functions
\begin{align}
C_\text{CDW}(i,j)=\langle \exp \(2 i \theta_i - 2 i \theta_j \) \rangle\\
C_\text{Nematic}(i,j)=\langle \exp \(2 i \phi_i - 2 i \phi_j \) \rangle
\end{align}
were measured, as well as the nematic stiffness parameter $\rho_N$, i.e. the response of the free energy $F$ to a twist $\Phi$ in the boundary conditions for $\phi$:
\begin{align}
\rho_N = \frac{1}{\kappa_s + 4 \kappa_c}\frac{\partial^2}{\partial \Phi^2}F(\Phi)\big|_{\Phi=0}, \label{eqn.rhon}
\end{align}
where in the absence of disorder, $\lim \limits_{T \rightarrow 0} \rho_s =1$. $\rho_N$ was obtained directly as a non-local correlations function on systems with periodic boundary conditions. The data is summarized in Fig. \ref{fig.bigfig}. 

As expected, equilibration becomes difficult to achieve at low temperatures due to slow glassly dynamics, putting severe constraints on the achievable system sizes. We ensured that equilibration is indeed achieved by using three different initializations and confirming that the measurements are independent of this choice. We took (i) disordered configuration with $\phi,\theta$ completely random, (ii) ordered configuration with $\langle e ^{i \theta}\rangle = \langle e ^{i \phi}\rangle \neq 0 $, (iii) an `annealing' protocol where the temperature is successively lowered towards the target value. 

\paragraph{Correlation functions}--  Typical results for the two correlations functions $C_\text{CDW}$ and $C_\text{Nematic}$ are shown in Fig. \ref{fig.bigfig} (left) for $\beta \kappa_s =10\beta v=5$, $\kappa_c=v$ measured for 15 disorder realizations on a system of size $160 \times 160$. For these system sizes, the spatial correlations exhibit very little sample dependence, and the combined error due to sampling and disorder averaging is smaller than the symbols used to plot the data. More generally, $C_\text{CDW}$ always exhibits rapid exponential decay at low temperatures and moderately weak disorder strength, while $C_\text{Nematic}$ decays much slower. Depending on the parameters $C_\text{Nematic}$ can also be exponential, or follow a power-law up to achievable system sizes.

\paragraph{Temperature dependent correlation length}--  At moderate temperatures, $C_\text{Nematic}$ is always follows an exponential decay with a temperature dependent correlations length $\xi_N$, shown in Fig. \ref{fig.bigfig} (center) for both $\kappa_c = -0.1 \kappa_s$ (SDW glass) and for $\kappa_c = 0.1 \kappa_s$ (fully disordered). At high temperatures $\xi_N(\kappa_c=0.1 \kappa_s) > \xi_N(\kappa_c=-0.1 \kappa_s)$, as would be the case for $v=0$ (no disorder). As temperature is lowered, disorder becomes important and affects the relative size of the $\xi_N$. For parameters corresponding to a SDW glass phase at lower temperatures, $\xi_N$ diverges. For parameters where single dislocations eventually proliferate, the correlation length shows no sign of divergence.

\paragraph{Helicity modulus}--  The stiffness parameter $\rho_N$ as defined in Eq.(\ref{eqn.rhon}) provides a sharp distinction between the two low temperature phase. A non-zero value for $\rho_N$ in the thermodynamic limit is a characteristic feature of the SDW glass phase while in the fully disordered phase where all vortices have proliferated $\lim \limits_{L \rightarrow \infty}\rho_N=0$. Results for the stiffness are shown in Fig. \ref{fig.bigfig} (right) for $\beta \kappa_s =10\beta v=5$, $\kappa_c=\pm v$. In the SDW phase, $\rho_N$ exhibits little variation between different samples and approaches a finite value at the largest system sized that we considered. In the phase where single dislocations proliferate, $\rho_N$ fluctuates strongly from sample to sample, around an average that rapidly decays with increasing system size.

Despite the limitations on sample size, the result of our simulations are fully consistent with the analytical predictions. In particular we find that depending on our choice of parameters, the system exhibits qualitatively different behavior, corresponding to two different phases at low temperatures---the SDW glass phase which we introduce here, and a more conventional glassy state where all correlations decay exponentially. We expect that these conclusions remain valid for $L \rightarrow \infty$ and $T \rightarrow 0$.

\section{Discussion}

Our results should be pertinent to a wide variety of systems. Below we highlight a few specific interesting examples. 

SDW ordering is very common in electronic solids, and is often incommensurate with the underlying lattice. The classic example\cite{chreview} is elemental Chromium $Cr$. There have been several studies of the suppression of SDW order in $Cr$ when it is alloyed with other transition metal elements (see Refs.~\onlinecite{rosenbaum,minhae} and references therein), for instance Vanadium $V$. At low $V$ concentrations where the disorder is weak, we expect our results to apply directly and predict the occurrence of the 3D SDW glass phase. Consequently the static spin structure factor is a power law (Eqn. \ref{3dsdwgsfc}), potentially visible in high resolution neutron diffraction studies. 

The SDW glass offers an interesting experimental opportunity to probe the physics of the 3D Bragg glass in magnetic systems. The original theoretical proposal\cite{GLD94,giam} of the Bragg glass phase spurred a search for it in a few experimental systems, notably in vortex matter inside superconductors (for a review see Ref.~\onlinecite{giambhatt}). Ref.~\onlinecite{klein} provided 
evidence for the predicted power law Bragg peaks in a disordered vortex lattice through small angle neutron scattering . Experimental evidence for Bragg glass physics in CDW systems seems scarce---probably due to the strong coupling to disorder of the CDW order parameter. Recently however Scanning Tunneling Microscopy images of the quasi-two dimensional CDW ordered system $NbSe_2$ have been interpreted\cite{pasupathy} in terms of a Bragg glass picture which might describe intermediate length scale physics. SDW systems of the kind considered in this paper 
offer a different context for Bragg glass physics which may be more directly amenable to experimental studies. 

We emphasize that the SDW glass is distinct from the conventional Heisenberg spin glass. The SDW glass is also distinct from the `cluster spin glass' which macroscopically is the same phase as the usual spin glass. It is interesting that even some `classic' metallic spin glasses\cite{mydosh,binderyoung} (for instance $CuMn$) actually have substantial short range SDW order\cite{mydosh} (visible in neutron diffraction as a well defined finite wave vector peak). Physically these are usefully understood as obtained from local pinning of SDW fluctuations of elemental $Cu$ around $Mn$ impurities. A useful theoretical approach to understanding these systems may be to start with the SDW glass described in this paper and then to disrupt it with topological defects at long length scales. 

Turning to $2d$ systems, it is interesting to consider the very lightly doped cuprates within the framework of our results. At low-$T$, these have long been reported to have spin glass order but also show substantial coexisting SDW correlations\cite{keimer}. As we have argued if the `parent' SDW order is uniaxial ({\em i.e} not a spiral) then two distinct kinds of glassy states are possible in both of which the SDW correlations decay exponentially on long scales. When only doubled dislocations of the accompanying CDW are induced (the 2D SDW glass phase), spin nematic order persists. If however single dislocations are also induced the resulting phase is smoothly connected to the conventional spin glass. 

What experiments can help distinguish between these two phases? As we discussed they will have rather similar peaks in neutron diffraction. More telling will be local probes of the spin dynamics for instance through NMR. The local dynamic spin susceptibility in the SDW glass phase should behave similarly to that in an ordered antiferromagnet (as the $\theta$ field is frozen the spin auto-correlation is determined entirely by $\vec N$), and will not show very striking glassy effects. In contrast there will be a wide range of relaxation times in a conventional spin glass. 

For PDW order the main proposed candidate to date are the cuprates, notably $La_{2-x}Ba_xCu_2O_4$. Close to $x = \frac{1}{8}$, there is an interesting window of temperatures between $4 K$ and $16 K$ where the in-plane resistivity is immeasurably small while there is a non-zero $c$-axis resistivity\cite{fktrev}. Further the Meissner effect itself onsets only below $4 K$. It has been suggested that this behavior may be explained by a PDW order pinned by impurities, and with frustrated $c$-axis Josephson coupling. Our results show that there are two possible fates of the PDW at weak disorder. In the PDW glass phase there are no frozen disorder-induced superconducting vortices. Consequently we expect a Meissner effect in this phase, and a non-zero critical current. In the fully disordered glass phase there are frozen random sign $\frac{hc}{4e}$ vortices. This phase is then likely to behave similarly to a vortex glass. It will presumably have vanishing linear resistivity but a zero critical current and no Meissner effect. Thus it has the potential\cite{bfk} to explain the experiments within the PDW framework. An immediate consequence is the local breaking of time reversal symmetry at zero field due to the frozen $\pm \frac{hc}{4e}$ vortices. It will be most interesting to look for this through scanning SQUID microscopy or other local probes of magnetism. 

Modulated superconductivity in the FFLO states, has of course, been discussed theoretically for decades. In recent years there have been suggestions of experimental sightings of this state in two different systems---first in the heavy fermion superconductor\cite{cecoid5-a,cecoid5-b} $CeCoIn_5$ and very recently in an organic superconductor\cite{Kanoda}. Both of these are very clean systems and hence our results on the effects of weak disorder may be directly applicable. $CeCoIn_5$ is a 3D superconductor and therefore may be in a superconducting Bragg glass phase. 
The organic is quasi-two dimensional and thus will at best be in a phase with long range charge-$4e$ superconducting order. In both materials it will be interesting to look for $\frac{hc}{4e}$ flux quantization.

We thank L. Balents, E. Fradkin, T. Giamarchi, D. Huse, and K. Moler for useful discussions and encouragement. DFM acknowledges support by the Caltech Institute for Quantum Information and Matter, an NSF Physics Frontiers Center with support of the Gordon and Betty Moore Foundation through Grant GBMF1250; and the Walter Burke Institute for Theoretical Physics at Caltech.
TS was supported by Department of Energy DESC-8739- ER46872, and partially by a Simons Investigator award from the Simons Foundation. 

\appendix
\section{Functional Renormalization Group for the 3D Larkin-Ovchinnikov superconducting glass}\label{app.fflofrg}
The starting point for the FRG analysis is the long-wavelength Hamiltonian
\begin{align}
 H_\text{FFLO} &= H_\text{CDW} + H_\text{SC} + H_\text{B}\\
 H_\text{CDW}& =K_c \int_{\vect x} \sum_n (\vect \nabla \theta_n)^2 + \beta \sum_{n,n
} V(\theta_n - \theta_{n'})\\
 H_\text{SC}& =K_s \int_{\vect x} \sum_n (\vect \nabla \phi_n)^2 \\
 H_\text{B}& =2 \lambda \int_{\vect x} \sum_n \vect \nabla \theta_n\cdot \vect \nabla \phi_n,
\end{align}
where $V(\theta)=V(-\theta)=V(\theta+ \pi)$ is a symmetric, periodic function and $\beta$ is the inverse temperature. In the SDW glass $\lambda=0$ and $\phi,\theta$ decouple. The renormalization group equations for $H_\text{CDW}$ were derived in Ref.~\onlinecite{fish} as
\begin{align}
&\frac{d}{d \ell} K_c =0\\
&\frac{d}{d \ell} V(\theta) = \epsilon V(\theta) + \frac{2V''(\theta)^2-4V''(\theta)V''(0)}{(2\pi)^4 K_c^2},
\end{align}
where $\epsilon = 4-d$ and the tree level scaling of the elastic term has been absorbed into $\beta$. In the present case of $\lambda \neq 0$, it is clear that $K_s$ and $\lambda$ cannot be renormalized by $V$. The only modification to the RG equations is $K_c \rightarrow \tilde K_c =K_c- \lambda^2/K_s$ which can be absorbed by a simple rescaling so long as $\lambda^2 < K_cK_s$. Thus the 3D FFLO glass exhibits the same universal properties as the 3D SDW glass discussed in Sect. \ref{sec.3d}.

\section{Alternative lattice model}\label{app.model}
In the main text we introduced the Hamiltonian
\small
\begin{align}
H = & -J\sum_{<ij>} \vec N_i \cdot \vec N_j \cos(\theta_i - \theta_j + \eta_{ij}) - v\sum_i \cos(2\theta_i - \alpha_i)\nonumber
\end{align}
\normalsize
to capture the low-energy properties of the order parameters, in particular the structure of topological defects and the coupling to the disorder potential. A key feature of this Hamiltonian is a large redundacy corresponding to local gauge invariance under $\vec N_i \rightarrow -\vec N_i$, $\theta_i \rightarrow \theta_i + \pi$. In some cases it is more convenient to adopt an alternative, equivalent formulation which makes this more explicit, and at the same time highlights the special role played by single dislocations. To this end we introduce a model in terms of $\vec N_i$ and $b_i = e^{i\theta_i}$ coupled to a $\mathbb Z_2$ gauge field $\sigma_{ij}$ with the Hamiltonian
\begin{eqnarray}
H_\text{alt} & = & -\sum_{ij}\sigma_{ij} \left( J_s \vec N_i \cdot \vec N_j + J_c \cos(\theta_i - \theta_j + \eta_{ij})\right) \nonumber \\
& & - v\sum_i \cos(2\theta_i - \alpha_i).
\end{eqnarray}
This has the same gauge invariance as $H$ provided that $\vec N_i \rightarrow -\vec N_i$, $\theta_i \rightarrow \theta_i + \pi$ is accompanied by $\sigma_{ij}\rightarrow -\sigma_{ij}$ for all sites $j$ connected to $i$ (the familiar `star' transformation in $\mathbb{Z}_2$ gauge theory). In the ground state one may choose the gauge $\sigma_{ij}=1$ to see that $H$ and $H_\text{alt}$ yield the same energy for smooth fluctuations of $\vec N$ and $e^{i \theta}$. Moreover, $H_\text{alt}$ clearly allows $2\pi$ vortices in $\vec N$ and double dislocations. 

In addition, there may be `visons' i.e. $\pi$ flux configurations in the $\sigma_{ij}$ which cannot be removed by a gauge transformation. This flux is seen by \emph{both} $\vec N$ and $e^{i \theta}$, and therefore induces a single dislocation tied to a half-vortex in $\vec N$. Thus the structure of topological defects in $H$ and $H_\text{alt}$ is identical. Since the divergent contribution to the energy of any allowed defect is determined by the elastic terms in the Hamiltonian, $H$ and $H_\text{alt}$ describe the same physics at long length scales.

\section{Estimates of length scales in 2D}\label{app.scales}

The presence or absence of isolated dislocations is determined by the balance
between elastic energy cost and energy gain due to disorder. The elastic energy
cost of a defect where $\theta$ winds by $2 \pi m_c$ and $\phi$ winds by $2 \pi
m_s$
\begin{align}
E_{m_c,m_s}= \pi(K_c m_c^2 + K_s m_s^2) \log L/a.
\end{align}
The energy gain $V_\text{dis}(\vect x)$ due to disorder depends on the position
$\vect x$ of the vortex and the particular disorder realization. The
distribution of $V_\text{dis}$ for a constant variance of the random force is
given by
\cite{natter}
\begin{align} 
&P_{m_c}(V_\text{dis})= \frac{1}{\sigma m_c \sqrt{2 \pi}}\exp\(-\frac{V_\text{dis}^2}{2
\sigma^2 m_c^2}\)\label{distribution}\\
&\sigma^2= 2 \pi D_F K_c^2 \log \frac{L}{a} 
\end{align}
In the presence of random-fields, the variance of the random force is itself
scale-dependent $D_F(L) \approx C \log \frac{L}{\xi_L}$ where $C$ is temperature
dependent and $\xi_L \gg a$ is the Larkin length. In this case, a reasonable
approximation consists of replacing the variance $D_F$ of the random
force by its average on a logarithmic scale
\begin{align}
D_F \rightarrow \bar D_F(L)\equiv \frac{1}{\log \frac{L}{\xi_L}}\int_{\xi_L} ^L\frac{D_F(R)}{R}dR \sim
\frac{C}{2}\log\frac{L}{\xi_L}
\end{align}
The probability $p(L)$ that it a vortex exists in a volume $L^2$ is given by
\begin{align}
p(L)&= \(\frac{L}{\xi_L}\)^2 \int_{-\infty}^{- E_{m_c,m_s}} P(V)\\
&\approx \frac{\sigma}{E_{m_c,m_s} \sqrt{2 \pi}}
\exp\(2 \log \frac{L}{\xi_L}-\frac{E_{m_c,m_s}^2}{2 m_c^2 \sigma^2}\). 
\end{align}
To estimate the scale where defects proliferate, we drop the subleading factor
and set $p(\xi_{m_c,m_s})=1$, obtaining 
\begin{align}
 \xi_{m_c,m_s} =\xi_L e^{\sqrt{\frac{\pi}{4 C m_c^2}( m_c^2 +
 K_s/K_c m_s^2)^2\log \xi_L/a}} 
\end{align}

\section{Long distance physics for large $\frac{K_s}{K_c}$ in $2d$}\label{app.gauge}
In this Appendix we sharpen the arguments justifying the absence of single dislocations at long wavelengths for large $\frac{K_s}{K_c}$ in 2D. This is most conveniently done within the alternative formulation of the model introduced in App. \ref{app.model}. Although $H_\text{alt}$ can be fruitfully used to discuss all length scales/parameter regimes, we will use it here as an effective model for the large $\frac{K_s}{K_c}$ regime at length scales longer than $\xi_V^{(2)}$ (the scale of doubled dislocation proliferation) discussed in the main text. Thus we will take the lattice spacing to be of order $\xi_V^{(2)}$. The long length scale physics is then captured by this model with $v \rightarrow \infty$. Note that the core energy of a single dislocation must be taken to be $E_\text{core} \sim K_c \ln\left(\frac{\xi_V^{(2)}}{a}\right)$. 

When $v \rightarrow \infty$ we have 
\begin{equation}
\theta_i - \frac{\alpha_i}{2} = \pi \frac{1- s_i}{2}
\end{equation}
with $s_i = \pm 1$. Then the Hamiltonian becomes
\begin{equation}
H_\text{eff} = -\sum_{ij}\sigma_{ij} \left( J_s \vec N_i \cdot \vec N_j + J_c s_i s_j \cos\left(\frac{\alpha_i - \alpha_j}{2} + \eta_{ij}\right) \right)
\end{equation} 
We define $\tilde{J}_{c, ij} = J_c \cos\left(\frac{\alpha_i - \alpha_j}{2} + \eta_{ij}\right)$. With $\eta_{ij}, \alpha_i$ random uncorrelated variables, $\tilde{J}_{c, ij}$ will have a probability distribution symmetrically distributed zero, and will be uncorrelated between different sites. In the absence of the $\sigma_{ij}$ the second term describes a 2D Ising spin glass. In the configuration of ${\tilde{J}_{c, ij}}$ there will with some probability be several frustrated plaquettes (where the product around the plaquette of $\tilde{J}_{c, ij}$ will be negative). If the gauge field $\sigma_{ij}$ adjusts itself to ``unfrustrate" those frustrated plaquettes the Ising subsystem will gain energy $\Delta E_\text{Ising} = \lambda J_c$ for some constant $\lambda$. 
However such a $\pi$-flux of $\sigma_{ij}$ nucleates a $\pi$ disinclination in $\vec N$ which costs energy $E_\text{core} \sim K_c \ln\left(\frac{\xi_V^{(2)}}{a}\right)$ in addition to the elastic energy of distorting the $\vec N$ upto the typical separation between two such frustrated plaquettes. Thus the cost of nucleating a $\pi$-disclination in $\vec N$ overwhelms any energy gain from unfrustrating the Ising spin $s_i$. Single dislocations are therefore suppressed in this regime.

\end{document}